\documentclass[12pt]{iopart}
\usepackage{cite}
\usepackage[dvips]{graphicx}
\renewcommand{\lsim}{~{\buildrel < \over {_\sim}}~}
\newcommand{\gsim}{~{\buildrel > \over {_\sim}}~}
\newcommand{\sqrtsNN}{\sqrt{s_{\scriptscriptstyle{{\rm NN}}}}}
\newcommand{\av}[1]{\left\langle #1 \right\rangle}

\newcommand{\mev}{\mathrm{MeV}}
\newcommand{\gev}{\mathrm{GeV}}
\newcommand{\tev}{\mathrm{TeV}}
\newcommand{\fm}{\mathrm{fm}}

\newcommand{\mum}{\mathrm{\mu m}}

\newcommand{\PbPb}{\mbox{Pb--Pb}}

\renewcommand{\AA}{\mbox{nucleus--nucleus}}
\newcommand{\RAA}{R_{\rm AA}}

\renewcommand{\pt}{p_{\rm t}}
\renewcommand{\d}{{\rm d}}

\newcommand{\ccbar}{\mbox{$\mathrm {c\overline{c}}$}}

\newcommand{\Dz}{\mbox{$\mathrm {D^0}$}}
\newcommand{\DtoKpi}{\mbox{${\rm D^0\to K^-\pi^+}$}}

\begin{document}

\title{Charm quenching in heavy-ion collisions at the LHC}

\author{Andrea Dainese\,\footnote{andrea.dainese@pd.infn.it}}

\address{Universit\`a degli Studi di Padova and INFN, 
         via Marzolo 8, 35131 Padova, Italy}

\begin{abstract}
D-meson suppression in Pb--Pb collisions at the LHC due to 
charm quark in-medium energy loss is estimated within a model that   
describes the available quenching measurements at RHIC. The result is compared
to that previously published by the author.
The expected sensitivity of the ALICE experiment for studying charm energy 
loss via fully-reconstructed $\Dz$-meson decays is also presented.
\end{abstract}




\section{Introduction}
\label{intro}

High-$\pt$ particle suppression in nucleus--nucleus (AA) with 
respect to pp collisions is regarded as one of the major discoveries at RHIC. 
The effect can be quantified with the nuclear modification factor: 
\begin{equation}      
   \label{eq:raa}
     R_{\rm AA}(\pt) \equiv 
     \frac{1}{\av{N_{\rm coll}}_{\rm centrality\,class}} \times
     \frac{{\rm d}^2 N_{\rm AA}/{\rm d}\pt{\rm d}\eta}
          {{\rm d}^2 N_{\rm pp}/{\rm d}\pt{\rm d}\eta} \,,  
\end{equation} 
which, at high $\pt$, 
would be equal to unity if the AA collision was a mere superposition 
of $N_{\rm coll}$ independent NN collisions. 
In central Au--Au collisions at centre-of-mass energy $\sqrtsNN=200~\gev$ per 
nucleon--nucleon (NN) pair the PHENIX and STAR experiments 
have measured $\RAA\simeq 0.2$
at high $\pt$ ($\gsim 4~\gev$) and central pseudorapidity 
($|\eta|\lsim 1$)~\cite{rhicRAA}.
The observed suppression is explained in terms of hard-parton energy loss 
(quenching) due to gluon radiation induced by the dense 
QCD medium expected to be formed in high-energy heavy-ion collisions 
(see~\cite{vitev,bdmps,qw} and references therein).
As the effect depends on the medium properties, and on its density in 
particular, hard partons can be used as probes to perform tomographic 
studies of the medium itself.

Charm and beauty quarks are
qualitatively different probes with respect to light partons, 
since the `dead-cone effect' is expected to reduce the in-medium gluon 
radiation off massive partons~\cite{dokshitzerkharzeev}. 
At the LHC, more than 100 $\ccbar$ pairs are expected to be produced per 
central \PbPb~collision at $\sqrtsNN=5.5~\tev$ and the detectors 
will be equipped with high position-resolution tracking systems, specifically
designed for heavy-flavour detection. 
Therefore, it will be important to carry out a comparative study of the 
quenching of massless and massive probes in order to (a) test the consistency 
of the interpretation of the effect as due to energy loss in a 
deconfined medium and (b) further investigate the properties (density) of 
such a medium.

\section{Energy loss for massive partons in the BDMPS formalism}
\label{eloss}

We calculate in-medium parton energy loss in the framework of the 
`BDMPS' (Baier-Dokshitzer-Mueller-Peign\'e-Schiff) formalism,
reviewed in~\cite{bdmps}. 

In a simplified picture, a parton produced in a hard collision 
undergoes, along its path in the medium, multiple scatterings
in a Brownian-like motion with mean free path $\lambda$. 
Consequently, the gluons in the parton wave function pick up
transverse momentum $k_{\rm t}$ with respect to its direction 
and they may eventually decohere and be radiated. 

The scale of the energy loss is set by the characteristic energy 
of the radiated gluons, $\omega_{\rm c} = \hat{q}\,L^2/2$,
which depends on the in-medium path length $L$ of the parton 
and on the BDMPS transport coefficient of the medium, $\hat{q}$. 
The transport coefficient is defined as the average 
medium-induced transverse momentum squared transferred to the parton 
per unit path length, 
$\hat{q} = \av{k_{\rm t}^2}_{\rm medium}\big/\lambda$~\cite{bdmps}.

In the case of a static medium, the distribution of the energy $\omega$ of the
radiated gluons (for $\omega\ll\omega_{\rm c}$) is of the form:
\begin{equation} 
\label{eq:wdIdw}
\omega\frac{{\rm d}I}{{\rm d}\omega}\simeq \frac{2\,\alpha_{\rm s}\,C_{\rm R}}{\pi}\sqrt{\frac{\omega_{\rm c}}{2\omega}}\,,
\end{equation}
where $C_{\rm R}$ is the QCD coupling factor (Casimir factor), equal 
to 4/3 for quark--gluon coupling and to 3 for gluon--gluon coupling.
In the case of infinite parton energy, $E\to\infty$, the integral of the 
radiated-gluon energy distribution up to $\omega_{\rm c}$ estimates the 
average energy loss of the parton:
\begin{equation}
\label{eq:avdE}
\av{\Delta E} = \int_0^{\omega_{\rm c}} \omega \frac{{\rm d}I}{{\rm d}\omega}{\rm d}\omega
\propto \alpha_{\rm s}\,C_{\rm R}\,\omega_{\rm c} \propto \alpha_{\rm s}\,C_{\rm R}\,\hat{q}\,L^2\,.
\end{equation}
The average energy loss is: proportional to $C_{\rm R}$ and, 
thus, larger 
by a factor $9/4$ for gluons than for quarks; proportional to the 
transport coefficient of the medium; proportional to $L^2$; independent of 
the parton initial energy $E$. However, when the realistic case of finite 
parton energies is considered, the energy loss $\Delta E$ has to be 
constrained to be smaller than $E$. As discussed in~\cite{D0epjc,pqm}, 
this effectively results in reducing the difference between quark and gluon 
average energy losses 
and in changing the $L$ dependence from quadratic to approximately linear. 
Moreover, since a consistent theoretical treatment of the finite-energy 
constraint is at present lacking in the BDMPS framework, approximations 
have to be adopted, thus introducing uncertainties in the results~\cite{pqm}.

Heavy quarks with moderate energy, i.e.  $m/E>0$, propagate with 
a velocity $\beta=\sqrt{1-(m/E)^2}$ significantly 
smaller than the velocity of light, $\beta=1$. 
As a consequence, in the vacuum, gluon radiation at angles $\Theta$ smaller 
than the ratio of their mass to their energy $\Theta_0=m/E$ is suppressed by 
destructive interference~\cite{dokshitzerdeadcone}. 
The relatively depopulated cone around the heavy-quark direction with 
$\Theta<\Theta_0$ is called `dead cone'.
In~\cite{dokshitzerkharzeev} it is argued that the dead-cone effect 
should characterize also in-medium gluon radiation and it is approximated by 
means of a suppression factor that multiplies the energy distribution of 
the radiated gluons (\ref{eq:wdIdw}):
\begin{equation}
  \label{eq:FHL}
  \frac{{\rm d}I}{{\rm d}\omega}\bigg|_{\rm Heavy}\bigg/
  \frac{{\rm d}I}{{\rm d}\omega}\bigg|_{\rm Light}=
  \left[1+\frac{\Theta_0^2}{\Theta^2}\right]^{-2}=
  \left[1+\left(\frac{m}{E}\right)^2\sqrt{\frac{\omega^3}{\hat{q}}}\right]^{-2}
  \equiv F_{\rm H/L}\,,
\end{equation} 
where the expression for the characteristic gluon emission 
angle~\cite{dokshitzerkharzeev} 
$\Theta\simeq (\hat{q}/\omega^3)^{1/4}$ has been used.
The heavy-to-light suppression factor $F_{\rm H/L}$ in (\ref{eq:FHL}) 
increases (less suppression) as the heavy-quark energy $E$ increases 
(the mass becomes negligible) and it decreases at large $\omega$, indicating 
that the high-energy part of the gluon radiation spectrum is drastically 
suppressed by the dead-cone effect.

A recent detailed calculation of the radiated-gluon energy 
distribution $\omega\,\d I/\d\omega$ in the case of massive 
partons~\cite{ursmassive} confirms the 
qualitative feature of lower energy loss for heavy quarks, although the 
effect is found to be quantitatively smaller than that derived with the 
dead-cone approximation of~\cite{dokshitzerkharzeev}. 
A comparison of the results obtained in the two cases 
for the D-meson suppression in central \PbPb~collisions at the LHC
will be presented at the end of section~\ref{estimate}. 

\section{Calculating nuclear modification factors}
\label{RAA}

\subsection{The model}
\label{model}

The nuclear modification factor $\RAA$ for D mesons is calculated in 
a Monte Carlo approach developed in~\cite{pqm} for 
light-flavour hadrons. In this approach 
partons are produced and hadronized with the PYTHIA event 
generator~\cite{pythia}
and energy loss is sampled combining the `BDMPS quenching weights' 
and a realistic description of the \AA~collision geometry.

The quenching weight is the parton-energy-loss probability distribution, 
$P(\Delta E)$, and it is calculated from the 
radiated-gluon energy spectrum $\omega\,\d I/\d\omega$~\cite{qw}.
Input parameters are: $\alpha_{\rm s}$, which we fix to 1/3; 
$C_{\rm R}$; $\hat{q}$ and $L$, 
which are different for every parton and depend on the medium geometry and 
density profile. 
The distribution of parton production points in the 
plane transverse to the beam direction and the density profile
of the medium are both parameterized with the product of the Glauber-model 
thickness functions of the two colliding nuclei. We do not consider 
the medium expansion, since it was shown~\cite{qw} that this effect
can be accounted for by using an equivalent static medium with a 
time-averaged transport coefficient $\hat{q}$. 
The procedure to compute, parton-by-parton, $\hat{q}$ and $L$ as integrals 
along the parton propagation direction in the transverse plane is 
described in~\cite{pqm}. The absolute magnitude of $\hat{q}$ is set by a 
$\sqrtsNN$-dependent scale parameter, $k$, and
the decrease of the medium density (and, hence, of $\hat{q}$) when going from 
central to peripheral collisions is automatically given by the decrease of 
the product of the thickness functions.

The Monte Carlo chain used to obtain the quenched hadron $\pt$ distribution
is:
\begin{enumerate}
\item Generation of a parton with PYTHIA (pp collisions).
\item Sampling of a production point and azimuthal propagation direction.
\item Determination of $\hat{q}$ and $L$ and, from these, of the
      $P(\Delta E)$ distribution.
\item Sampling of a $\Delta E$ to be subtracted from the initial parton 
      energy $E$. As aforementioned, the BDMPS calculations are done in 
      the limit $E\to\infty$; therefore, for small $E$, part of the
      $P(\Delta E)$ distribution lies above $E$. Since there is no unique 
      way to implement the finite-energy constraint $\Delta E\le E$,
      we use two approaches~\cite{pqm}  
      and we use the band between the two results to visualize the
      theoretical uncertainty. 
\item Hadronization of the parton to a hadron according to a fragmentation 
      function.
\end{enumerate} 
The nuclear modification factor is calculated as the ratio of the quenched 
to unquenched (only steps (i) and (v)) $\pt$ distributions.

\subsection{Light-flavour hadrons suppression from RHIC to LHC}
\label{RAAh}

\begin{figure}[t!]
  \begin{center}
    \includegraphics[width=0.49\textwidth]{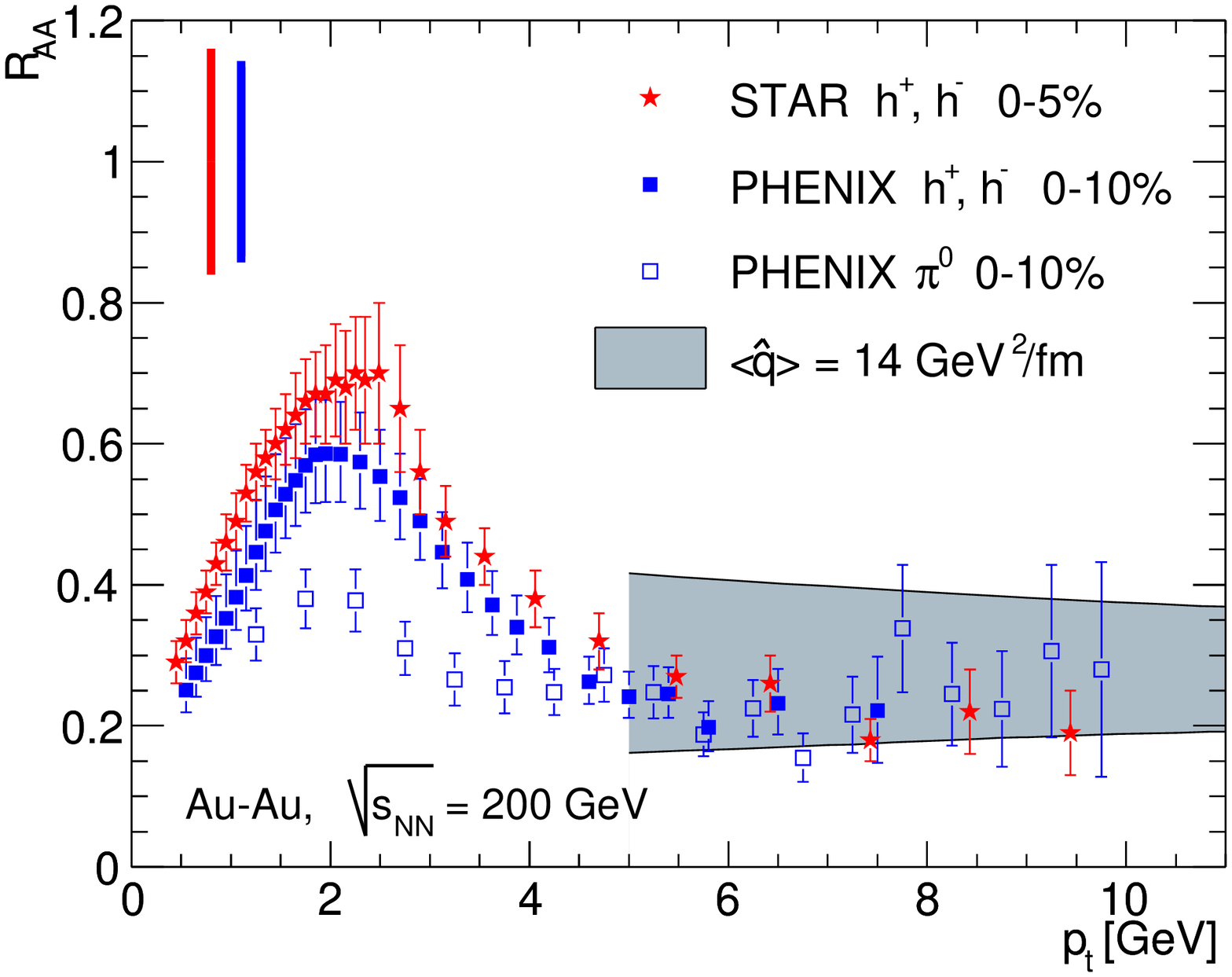}
    \includegraphics[width=0.49\textwidth]{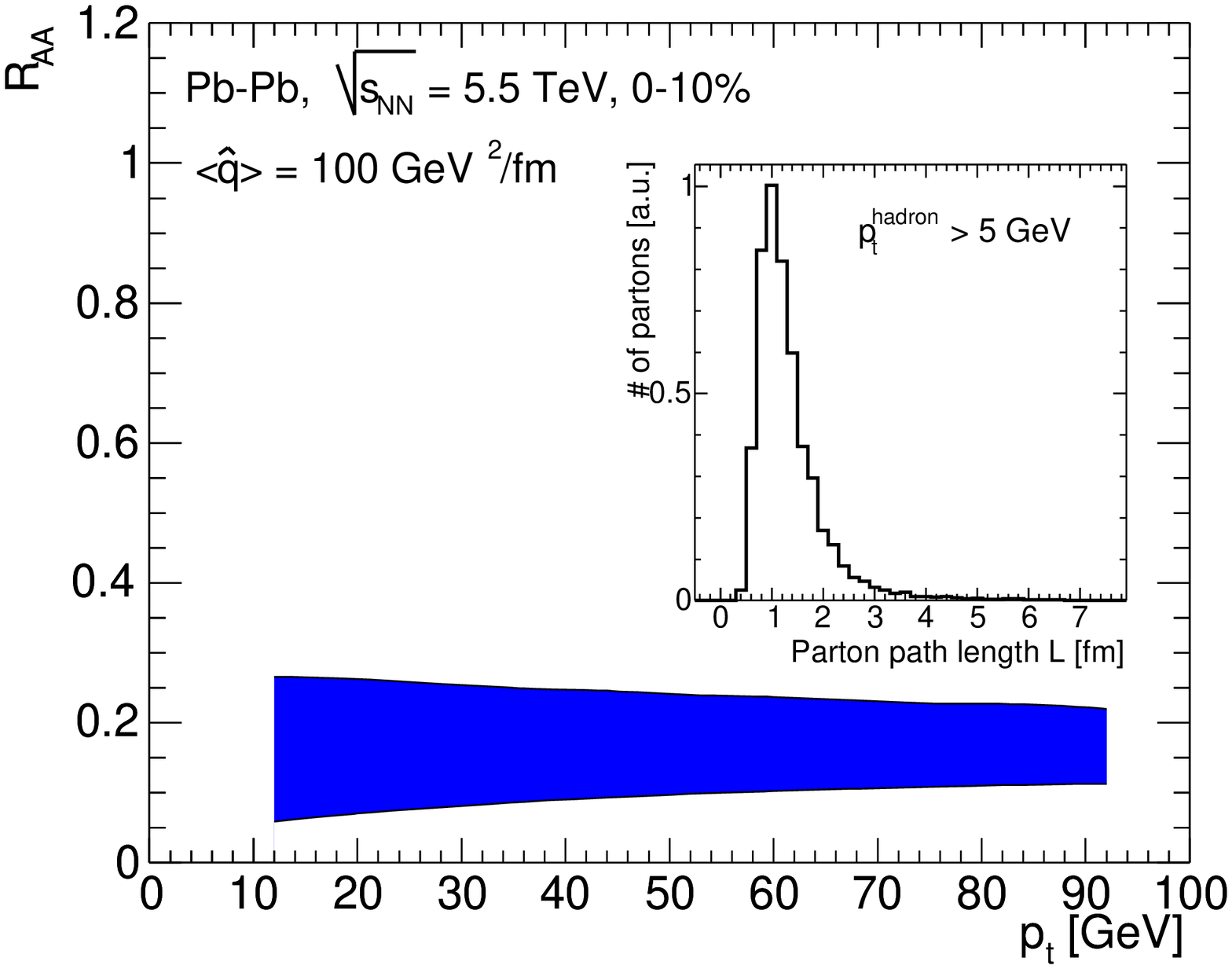}
    \caption{Charged hadrons ($h^\pm$) 
             and $\pi^0$ $\RAA(\pt)$ in central collisions
             at RHIC (left) and LHC (right). PHENIX 
             and STAR data~\cite{rhicRAA} are reported with combined
             statistical and $\pt$-dependent systematic errors (bars on the 
             data points) and $\pt$-independent systematic errors (bars at 
             $\RAA=1$). The model bands correspond to 
             $\av{\hat{q}}\approx 14~\gev^2/\fm$ (RHIC) and 
             $\av{\hat{q}}\approx 100~\gev^2/\fm$ (LHC)~\cite{pqm}.
             For the LHC case, the inset shows the $L$ distribution  
             for partons that escape the medium and fragment into hadrons 
             with $\pt>5~\gev$.} 
    \label{fig:RAAh}
  \end{center}
\end{figure}

The model was tuned for light-flavour hadrons 
in central \mbox{Au--Au} collisions at 
$\sqrtsNN=200~\gev$~\cite{pqm}. 
In this case PYTHIA was used with CTEQ4L parton distribution 
functions~\cite{cteq4} 
to generate light quarks (u, d) and gluons and 
hadronization was done with KKP fragmentation functions~\cite{kkp}. 
The scale parameter $k$ was chosen in order to 
reproduce the $\RAA$ suppression measured by PHENIX and STAR for 
charged hadrons ($h^\pm$) and $\pi^0$. 
The result is shown in the left-hand panel 
of Fig.~\ref{fig:RAAh}: the $\pt$-independence of the suppression for 
$\pt\gsim 5~\gev$\footnote{Since, for light-flavour hadrons, 
we did not include the nuclear modification of the parton distribution
functions, we show the model results for $\pt>5~\gev$ at RHIC 
energy and for $\pt>10~\gev$ at LHC energy, where this effect is 
expected to be small.} is well reproduced and 
the ensuing parton-averaged $\av{\hat{q}}$ is $\approx 14~\gev^2/\fm$.
The measured centrality dependence of the $\RAA$ suppression, as well as 
that of the back-to-back jet-like di-hadron correlations disappearance, 
is correctly reproduced by the model without changing the value of 
$k$~\cite{pqm}.

The extrapolation of $\hat{q}$ from RHIC to LHC energy was done by assuming 
the scaling $\hat{q}\propto n^{\rm gluons}$, where $n^{\rm gluons}$ is the 
initial volume-density of gluons, which, according to the saturation 
model~\cite{ekrt}, increases by a factor $\approx 7$ when $\sqrtsNN$ increases
from $200~\gev$ to $5.5~\tev$. 
The simulation for charged hadrons at the LHC was performed by running PYTHIA
at $\sqrtsNN=5.5~\tev$ and computing energy loss with  
$k_{\rm LHC}=7\, k_{\rm RHIC}$ 
(i.e. $\av{\hat{q}}_{\rm LHC}\approx 100~\gev^2/\fm$)~\cite{pqm}. 
The resulting 
nuclear modification factor in central \PbPb~collisions, shown in the 
right-hand panel of Fig.~\ref{fig:RAAh}, is of $0.1$--$0.2$ and rather 
independent of $\pt$ up to $100~\gev$. The path-length distribution 
for partons that escape the 
medium with large enough energy to fragment into hadrons with $\pt>5~\gev$,
shown in the inset, reveals that, within this model extrapolation, 
only partons produced close to the surface ($L\lsim 2~\fm$) can escape the 
medium, while all other partons are absorbed. 

\subsection{Estimate of D-meson suppression at the LHC}
\label{estimate}

\begin{figure}[t!]
  \begin{center}
    \includegraphics[width=0.7\textwidth]{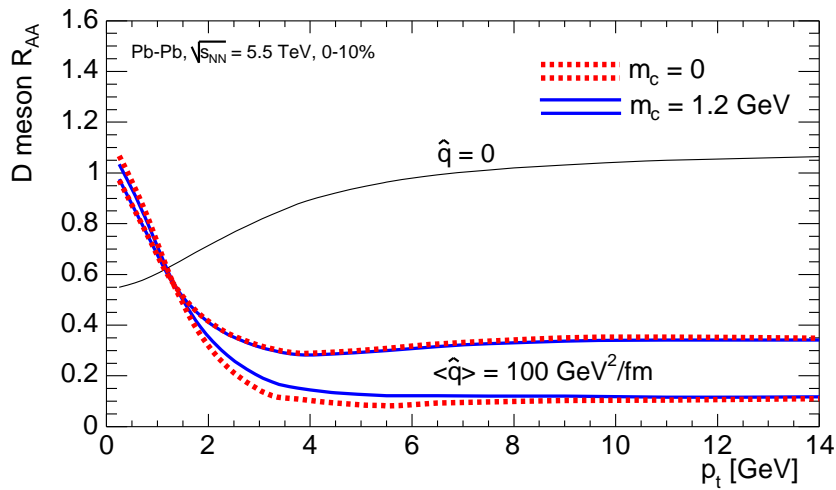}
    \caption{D-meson nuclear modification factor in central (0--10\%) 
             \PbPb~collisions at the LHC~\cite{adsw}. The curve labelled 
             ``$\hat{q}=0$'' includes only nuclear shadowing. 
             The two bands include {\it also} energy loss for massless 
             (dashed) or massive (solid) c quarks, with a 
             quark-averaged $\av{\hat{q}}\approx 100~\gev^2/\fm$.} 
    \label{fig:RAAD1}
  \end{center}
\end{figure}

In order to study the nuclear modification of the D mesons 
in central \PbPb~collisions at the LHC, we generate c quarks with PYTHIA
using a set of parameters that allows to reproduce the $\pt$ distribution 
given by next-to-leading order (NLO) perturbative QCD (pQCD) 
calculations~\cite{hvqmnr} with 
charm mass $m_{\rm c}=1.2~\gev$ and factorization and renormalization 
scales $\mu_F=\mu_R=2\,m_{\rm t}$, where $m_{\rm t}=\sqrt{\pt^2+m_{\rm c}^2}$ 
is the c-quark 
transverse mass, (see~\cite{noteHVQ}). We use CTEQ4L
parton distribution functions (PDFs) and we include, for \PbPb, their nuclear 
modification (shadowing) in the EKS98 parameterization~\cite{eks}.
For the hadronization to D mesons, we use a fragmentation function
parameterized from PYTHIA pp simulations done with the standard Lund string 
model.

The energy loss is sampled from the quenching weights for massive partons
calculated~\cite{adsw} with the full massive formalism developed 
in~\cite{ursmassive}. The sampling procedure is the same as for charged 
hadrons (section~\ref{RAAh}) and the same scale parameter 
$k_{\rm LHC}=7\,k_{\rm RHIC}$ is used. Differently from the case of 
charged hadrons, c quarks that lose all their energy, $\Delta E=E$, 
are `thermalized' and assigned a transverse momentum according to 
the transverse mass distribution 
$\d N/\d m_{\rm t}\propto m_{\rm t}\exp[-m_{\rm t}/(300~\mev)]$;
thus, the total c-quark production cross section is conserved.

Figure~\ref{fig:RAAD1} shows the estimated $\RAA$ for D mesons. The effect 
of shadowing is visible as a suppression at low $\pt$
in the curve without energy loss. The EKS98 parameterization
gives a reduction of a factor about 0.65 for the total $\rm c\overline c$ 
cross section and the suppression, at the D-meson level, is limited to the 
region $\pt\lsim 7~\gev$, corresponding to partonic momentum fractions
($x$ Bjorken) $x_1\simeq x_2\lsim 3\times 10^{-3}$.
Since there is a significant uncertainty on the magnitude of shadowing in 
this $x$ region, we studied the effect of such
uncertainty on $\RAA$ by varying the modification of the PDFs in a 
Pb nucleus. Even in the case of shadowing 50\% stronger than in EKS98, we 
found $\RAA>0.93$ for $\pt>7~\gev$~\cite{thesis}. We can, thus, conclude 
that c-quark energy loss can be cleanly studied, being the only expected 
effect, for $\pt\gsim 7~\gev$.

The results that include shadowing {\it and} energy loss, with 
$\av{\hat{q}}\approx 100~\gev^2/\fm$, are presented as two 
bands, obtained with $m_{\rm c}=0$ and $m_{\rm c}=1.2~\gev$~\cite{adsw}. 
According to this estimate and within the present theoretical uncertainties, 
the two results, massless and massive, do not differ significantly. The 
predicted D-meson $\RAA$ is in the range 0.1--0.35 and essentially 
$\pt$-independent for $\pt>7~\gev$.

In a previous estimate~\cite{thesis,D0epjc}, 
done before the analysis of RHIC data 
that we have summarized in section~\ref{RAAh}, we had employed a similar Monte 
Carlo procedure and description of the system geometry but a much smaller 
value for the transport coefficient, $\hat{q}=4~\gev^2/\fm$, that was expected 
to be `reasonable' for the LHC (see e.g.~\cite{baierQM02}). As the full massive
calculation of the radiated-gluon energy distribution~\cite{ursmassive} was 
not yet available, we had included the effect of the c-quark mass by 
means of the dead-cone approximation (\ref{eq:FHL}) 
of~\cite{dokshitzerkharzeev}. In that work we had found a significant 
difference between the massive and massless results~\cite{thesis,D0epjc}.
In Fig.~\ref{fig:RAAD2}, left-hand panel, we show how the results
of~\cite{thesis,D0epjc} with $\hat{q}=4~\gev^2/\fm$ and the dead cone evolved 
into the present ones with $\hat{q}=100~\gev^2/\fm$ and the full massive 
calculation (for simplicity,
we plot only the curves corresponding to the lower bound of the 
uncertainty bands of Fig.~\ref{fig:RAAD1}). As a first step, keeping 
$\hat{q}=4~\gev^2/\fm$, we remove 
the dead-cone approximation (curve ``a'') and use the full massive 
calculation (curve ``b''): the difference with 
respect to the massless result (curve ``c'')  
is already reduced. Then, we increase $\hat{q}$ to $100~\gev^2/\fm$: 
the suppression is much stronger and the difference between massive 
(curve ``d'') and massless (curve ``e'') is
further reduced. We are now in a scenario in which only partons produced 
nearby the surface escape and the (small) difference induced by the 
mass of the c quark on the gluon-radiation phase space becomes a minor effect.
The effect should be more significant for the heavier b quarks~\cite{adsw}.  

\section{Measurement of the D-meson nuclear modification factor with ALICE}
\label{alice}

\begin{figure}[t!]
  \begin{center}
    \includegraphics[width=0.49\textwidth]{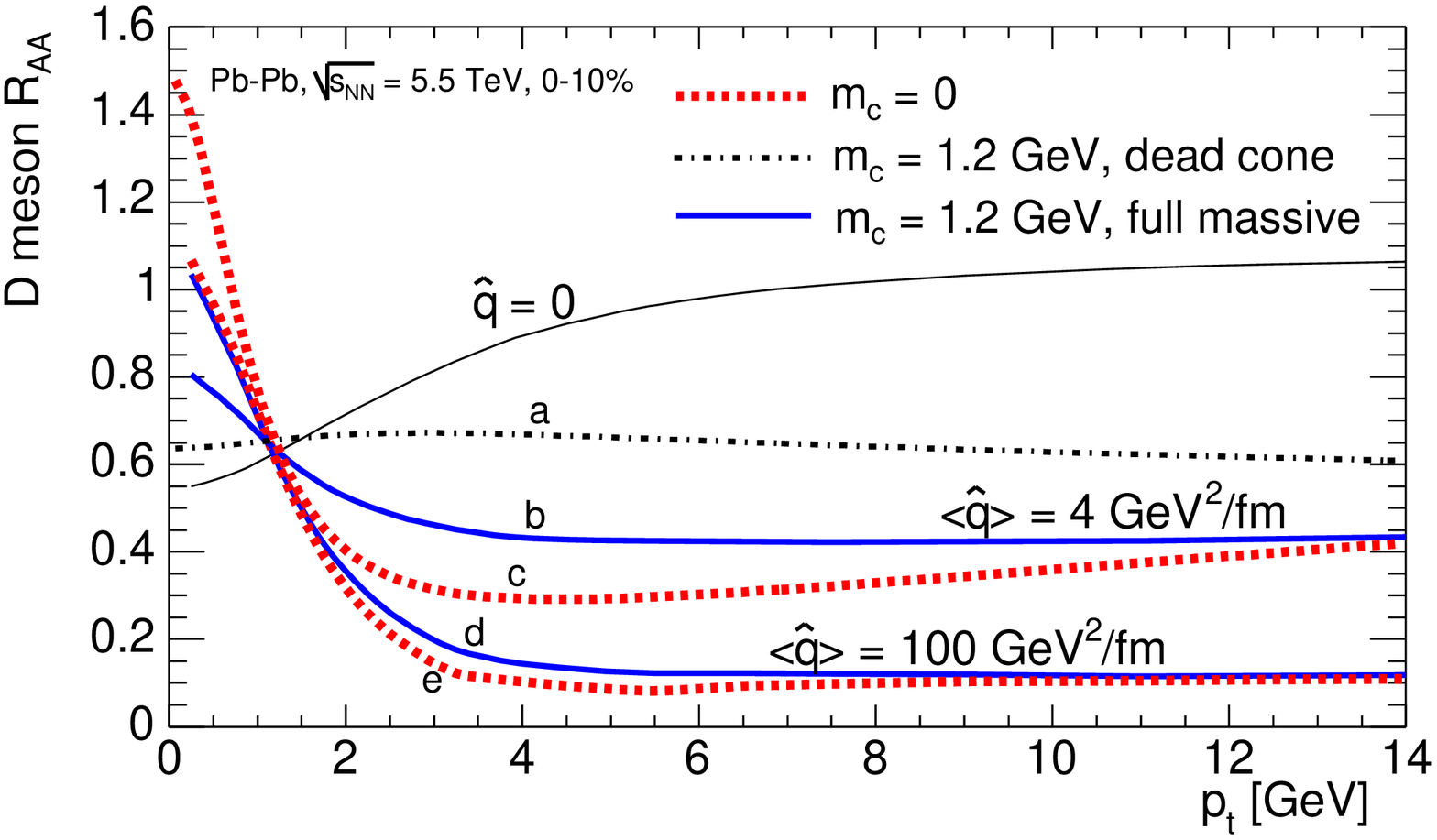}
    \includegraphics[width=0.49\textwidth]{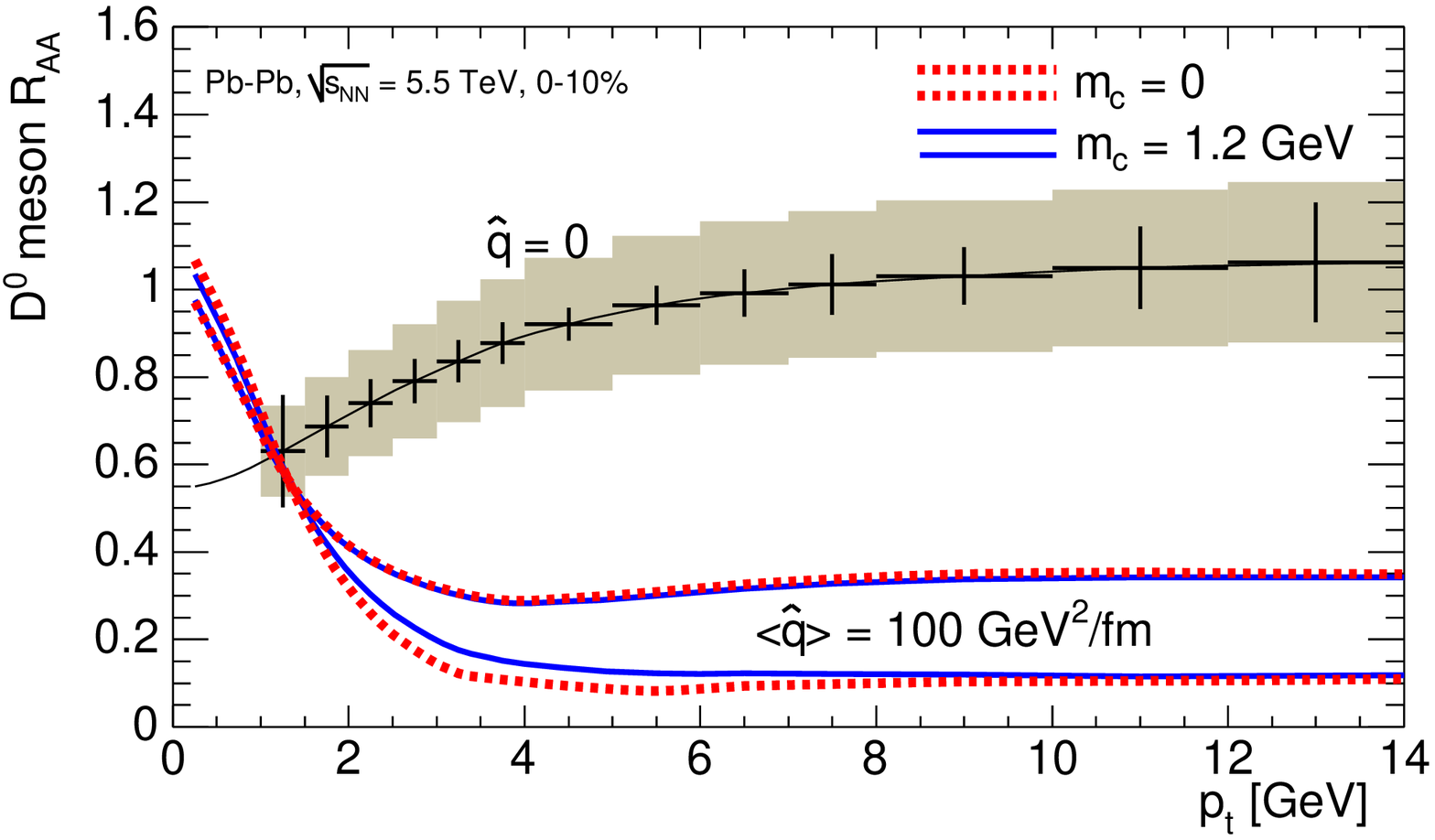}
    \caption{Left: current results on D-meson $\RAA$ compared to those 
             previously published in~\cite{D0epjc} (see text for details).
             Right: expected sensitivity attained by ALICE with 
             $\DtoKpi$ reconstruction; bars = statistical errors, 
             shaded area = combined systematic errors (see text).} 
    \label{fig:RAAD2}
  \end{center}
\end{figure}

The transverse momentum distribution of $\Dz$ mesons produced at central 
rapidity, $|y|<1$, can be directly measured from the exclusive 
reconstruction of $\rm D^0\to K^-\pi^+$ decays (and charge conjugates) 
in the Inner Tracking System (ITS), Time Projection Chamber (TPC) and 
Time Of Flight (TOF) detectors of the ALICE barrel, 
$|\eta|<0.9$~\cite{alicePPR}. 
The expected production yields per unit of rapidity at central rapidity 
for $\rm D^0$ (and $\rm \overline{D^0}$) mesons decaying in a 
$\rm K^\mp\pi^\pm$ pair,
estimated~\cite{noteHVQ} on the basis of NLO pQCD calculations with 
$m_{\rm c}=1.2~\gev$ and $\mu_F=\mu_R=2\,m_{\rm t}$, 
are $BR\times\d N/\d y=5.3\times 10^{-1}$ in central (0--5\%) 
Pb--Pb collisions at $\sqrtsNN=5.5~{\rm TeV}$ and 
$BR\times\d N/\d y=7.5\times 10^{-4}$ in pp collisions at 
$\sqrt{s}=14~{\rm TeV}$.
The main feature of the $\Dz$ decay topology is the presence of two tracks
displaced from the interaction point by, on average, $50~\mum$, for 
D-meson $\pt\simeq 0.5~\gev$, to $120~\mum$, for $\pt\gsim 5~\gev$.
Such displacement can be resolved with the ALICE tracking detectors
and thus a large fraction of the combinatorial background in the 
$\rm K^\mp \pi^\pm$ invariant mass distribution can be rejected.
The low value of the magnetic field, 0.4~T, and the $\rm K/\pi$ separation in 
the TOF detector extend the $\Dz$ measurement down to $\pt\approx 1~\gev$. 
The analysis strategy and the pertinent selection cuts were studied with a 
detailed simulation of the detector geometry and response, including the main
background sources~\cite{thesis,D0jpg}. 
The accessible $\pt$ range is $1$--$14~\gev$ for \PbPb~and 
$0.5$--$14~\gev$ for pp collisions.
The statistical error corresponding to 1 month of \PbPb~data-taking 
($\sim 10^7$ central events) and 9 months of pp data-taking ($\sim 10^9$ 
events) is better than 15--20\% and the systematic error 
(acceptance and efficiency corrections, subtraction of the feed-down from 
${\rm B}\to \Dz+X$ decays, cross-section normalization, 
centrality selection for \PbPb) is better than 20\%~\cite{thesis}.

In Fig.~\ref{fig:RAAD2} (right-hand panel) we show the expected sensitivity 
for the measurement of the $\Dz$ meson $\RAA$.
The reported errors, associated to the curve with shadowing and {\it no}
energy loss, are obtained combining the previously-mentioned errors 
in \PbPb~and in pp collisions and considering that several systematic 
contributions will partially cancel out in the ratio.
The uncertainty of about 5\% introduced in the extrapolation of the pp 
results from 14~TeV to 5.5~TeV by means pQCD, estimated in~\cite{thesis}, 
is also included. In the $\pt$ range $7$--$14~\gev$, where energy loss 
is the only expected effect, $\RAA$ can be measured with systematic errors
better than 20\% and statistical errors better than 15\%.

\section{Conclusions}
\label{concl}

We estimated the nuclear modification factor for
open-charm mesons in central \PbPb~collisions at LHC energy,
including PDFs shadowing and parton energy loss effects.
The latter were simulated using a model that combines the BDMPS 
quenching weights (specifically calculated for heavy quarks) 
and a realistic description of 
the collision geometry. The medium transport coefficient was extrapolated, 
according to 
the saturation model, from the value needed to describe $h^\pm$ 
and $\pi^0$ data at RHIC.

The D-meson suppression due to energy loss is found to be of approximately
a factor of 5 for $\pt\gsim 4~\gev$. The comparison of our results 
with $m_{\rm c}=0$ and $1.2~\gev$ 
suggests that the c-quark mass may not
reduce significantly the suppression, in contrast to what expected 
from the dead-cone effect. This is due to 
(a) the fact that the dead-cone effect resulting from the full massive
calculation is indeed quite small for charm quarks and 
(b) the fact that, within this model, the medium is found to be very opaque, 
already at RHIC energies, and only partons produced nearby the surface can 
escape.
Note, however, that the results have a significant uncertainty, because
a rigorous theoretical treatment of the finite parton energies is 
still lacking in all parton energy loss calculations.

The ALICE experiment, whose inner tracker is specifically optimized for 
heavy-flavour measurements, can address the intriguing phenomenology of 
charm quenching by reconstructing $\DtoKpi$ decays and measuring 
with good sensitivity the D nuclear modification factor up to about
15~GeV in transverse momentum.

\paragraph{Acknowledgement.}
The author, member of the ALICE Collaboration, acknowledges the 
ALICE off-line group, within which part of this work was carried out.

\vspace{.5cm}


\begin{thebibliography}{99}

\bibitem{rhicRAA} 
  Klay J {\it these proceedings} {\it Preprint} nucl-ex/0410033

\bibitem{vitev} 
  Vitev I {\it these proceedings} {\it Preprint} hep-ph/0409297

\bibitem{bdmps}
  Baier R, Schiff D and Zakharov B G 2000 {\it Ann.~Rev.~Nucl.~Part.~Sci.}
  {\bf 50} 37 

\bibitem{qw}
  Salgado C A and Wiedemann U A 2003 {\it Phys.~Rev.}~D~{\bf 68} 014008

\bibitem{dokshitzerkharzeev}
  Dokshitzer Yu L and Kharzeev D E 2001 {\it Phys.~Lett.}~B~{\bf 519} 199

\bibitem{pqm}
  Dainese A, Loizides C and Paic G 2005 {\it Eur.~Phys.~J.}~C~{\bf 38} 461

\bibitem{dokshitzerdeadcone}
  Dokshitzer Yu L, Khoze V A and Troyan S I 1991 {\it J.~Phys.}~G~{\bf 17} 1602

\bibitem{ursmassive}
  Armesto N, Salgado C A and Wiedemann U A 2004 {\it Phys.~Rev.}~D~{\bf 69} 
  114003

\bibitem{pythia} 
  Sj\"ostrand T {\it et al.} 2001 {\it Computer Phys. Commun.} {\bf 135} 238

\bibitem{cteq4}
  Lai H L~{\it et al.} 1997 {\it Phys.~Rev.}~D~{\bf 55} 1280

\bibitem{kkp}
  Kniehl B A, Kramer G and P\"otter B 2000 {\it Nucl. Phys.}~B~{\bf 582} 514

\bibitem{ekrt}
  Eskola K J, Kajantie K, Ruuskanen P V and Tuominen K 
  2000 {\it Nucl.~Phys.}~B~{\bf 570} 379

\bibitem{hvqmnr} 
  Mangano M, Nason P and Ridolfi G 1992 {\it Nucl.~Phys.}~B~{\bf 373} 295

\bibitem{noteHVQ}
  Carrer N and Dainese A 2003 ALICE-INT-2003-019 {\it Preprint} hep-ph/0311225 

\bibitem{eks} 
  Eskola K J, Kolhinen V J and Salgado C A 1999 {\it Eur.~Phys.~J.}~C~{\bf 9} 
  61

\bibitem{adsw}
  Armesto N, Dainese A, Salgado C A and Wiedemann U A
  {\it Preprint} hep-ph/0501225

\bibitem{D0epjc}
  Dainese A 2004 {\it Eur.~Phys.~J.}~C~{\bf 33} 495

\bibitem{thesis}
  Dainese A 2003 Ph.D. Thesis Universit\`a degli Studi di Padova {\it Preprint} nucl-ex/0311004

\bibitem{baierQM02}
  Baier R 2003 {\it Nucl.~Phys.}~A~{\bf 715} 209

\bibitem{alicePPR}
  ALICE Physics Performance Report Vol.~I (CERN/LHCC 2003-049) 2004  
  {\it J.~Phys.}~G~{\bf 30} 1517

\bibitem{D0jpg}
  Carrer N, Dainese A and Turrisi R 2003 {\it J.~Phys.}~G~{\bf 29} 575

\end{thebibliography}
\end{document}